# Slicing at the Physical Layer

Ana Pérez-Neira, *Fellow, IEEE*, Miguel Angel Lagunas, *Fellow, IEEE*

*Abstract*— **In next generation communications, slicing enables the selection and allocation of network resources to suit the requirements of very different vertical-driven use cases and applications. This work addresses fast real-time resource slicing by proposing the novel concept of slicing at the physical layer. To implement such a concept an orthonormal transform is devised that splits the Orthogonal Frequency Division Multiplex (OFDM) symbol into slices with ranked rate and latency. The advantage over any carrier or time segmentation of the original frame is that the proposed technique does not require channel information at transmission. Also, the frame overhead improves, and the decoding time is kept simple.**

## I. INTRODUCTION

The so-called network slicing has been identified as a key function for 5th Generation New Radio (5G NR) networks. The slices allow to share the same network infrastructure resource (e.g., spectrum, base stations, etc.) among multiple mobile network operators. For each slice, the infrastructure provider specifies the amount of network resources that the operators can use to serve the multitude of tenants [1]. In fact, the slice concept enables the upcoming shift from infrastructure planning to continuous (and likely dynamically adapting) real-time sharing and operation of the network. The infrastructure will not be under-utilized as it was with static and exclusive allocation policies. However, if network slicing wants to succeed in the above sense, the service creation time is required to be radically low and all the employed technologies must embrace this change. For this reason, this article dives into the possibility of including slices implemented at the hardware level of the physical layer. Existing works have addressed slicing at different levels of network abstractions, but the physical layer has so far been forgotten. Only recently [2] has brought out the importance of designing algorithms for a fine-grained partition of the network resources.

In this work, the background of slices in the physical layer is motivated by the formalism used for polar codes, [3], and the concept of channel polarization. In fact, it can be said that the channel is polarized in order to accommodate in a single hardware platform several physical channels with different rate and complexity/latency. In other words, polarization is understood as transforming the physical temporal channel into channels with controlled reliability.

We view that the notion of slices is embedded in the Discrete Fourier Transform/Inverse Discrete Fourier Transform (DFT/IDFT), like in polar codes butterfly construction. Therefore, the polarization of a given channel in several sub-channels is behind the well-known OFDM. However, the IDFT/DFT as polarizer suffers from a severe problem because the high- and low-quality flat fading channels are not provided in a ranked manner. In other words, they just follow the frequency response of the global channel. This differs from the use of Singular Value Decomposition (SVD), where the eigenvalues are ranked, [4-5]. The price to pay is that perfect Channel State Information at the Transmitter (CSIT) is required to implement the SVD, together with a higher complexity than the IDFT/DFT. This work presents a transform that, applied to the OFDM symbol, overcomes the IDFT/DFT mentioned drawbacks and converts the channel into channels with ranked rate and latency.

The novel scheme in this paper generates slices at the physical layer, where rate and complexity scale up and down without requiring CSIT. The only requirement is that the transmitter frames the symbol stream into their corresponding slice so that they can be properly accessed at the receiver. Each slice is tailored to provide a service with a specific requirement of quality of service and experience, as for instance, video content delivery or machine type communication. Then, all the generated physical layer slices are packed in one OFDM symbol, and, therefore, only one cyclic prefix (CP) is needed to transmit all the slices. In this way, the system gains in efficiency and simplicity with respect to the conventional or straight forward approach of devoting different OFDM symbols for each of the slices.

The following section introduces the basic splitting of the channel and the transformation to be applied to an OFDM symbol in order to create different physical slices. Next, section III computes the Quality of Service (QoS) in terms of the mutual information that each slice provides. Section IV presents possible receivers and motivates a specific transform design in order to keep on using the simplicity of the Fast Fourier Transform (FFT) algorithm at reception. Section V presents in detail the general recursive architecture for the physical layer slicing, and section VI supports its benefits with numerical simulations. Finally, section VII concludes.

*Notation*: Upper- and lower-case boldface letters denote matrices and vectors, respectively. Let the superscripts $(.)^T$ and $(.)^H$ denote transpose and Hermitian transpose operations, respectively. By $\boldsymbol{I}_N$ we denote the $N$-th order identity matrix, diag($\boldsymbol{a}$) is a diagonal matrix with the entries of the vector $\boldsymbol{a}$ on its diagonal. Finally, let $\sqrt{-1}$ be denoted by $j$.

## II.  THE BASIC CHANNEL SPLITTING

OFDM is a polarization scheme where the time dispersive (i.e., frequency selective) channel is transformed into $N$ flat narrow band channels. In order to confer this polarization with desirable properties for physical layer slicing, we propose to incorporate a basic channel splitting by applying the following orthonormal transformation

$$T_N = \frac{1}{\sqrt{2}}\begin{pmatrix} I_{N/2} & W_{N/2} \\ I_{N/2} & -W_{N/2} \end{pmatrix} \tag{1}$$

where $N = 2^k$ for $k$ integer, matrix $W \in C^{N/2 \times N/2}$ is orthonormal. Note that

$$T_2 = \frac{1}{\sqrt{2}}\begin{pmatrix} 1 & 1 \\ 1 & -1 \end{pmatrix} \tag{2}$$

and that $T_1 = 1$.

At the transmitter, this transformation is applied in baseband to the OFDM symbols, after the IDFT and before the insertion of the CP, which is longer than the channel length, $L$. At the receiver, once the CP is removed, the samples at base band can be collected in vector $y$, which is modeled as

$$y = H_{ZN} T_N s + w \tag{3}$$

where $w \in C^{N \times 1}$ is the additive white Gaussian noise of power equal to $\sigma^2$, and $s \in C^{N \times 1}$ collects the OFDM time samples after the IDFT. $H_{ZN} \in C^{N \times N}$ is the effective time channel matrix, which is circulant. Whenever the channel is of length $L \leq N/2$, this matrix can be formulated as

$$H_{ZN} = \begin{pmatrix} H_{N/2} & H_{CN/2} \\ H_{CN/2} & H_{N/2} \end{pmatrix} \tag{4}$$

where $H_{N/2} \in C^{N/2 \times N/2}$ is the following lower triangular and Toepltiz matrix

$$H_{N/2} = \begin{pmatrix} h_o & 0 & \cdots & 0 \\ h_1 & h_o & \cdots & 0 \\ \cdots & \cdots & \cdots & \cdots \\ 0 & h_{L-1} & \cdots & h_o \end{pmatrix} \tag{5}$$

with the samples $h_n, n = 0, \ldots, L-1$, denote the channel impulse response (CIR), and $H_{CN/2} \in C^{N/2 \times N/2}$ is the circular complement to $H_{N/2}$, which is due to the CP,

$$\boldsymbol{H}_{CN/2} = \begin{pmatrix} 0 & \cdots & h_{L-1} & \cdots & h_2 & h_1 \\ 0 & \cdots & 0 & h_{L-1} & \cdots & h_2 \\ \cdots & \cdots & \cdots & 0 & h_{L-1} & \cdots \\ \cdots & \cdots & \cdots & \cdots & 0 & h_{L-1} \\ \cdots & \cdots & \cdots & \cdots & \cdots & 0 \\ 0 & \cdots & 0 & \cdots & 0 & 0 \end{pmatrix} \quad (6)$$

Note that for $L \le N/2$ this circular complement matrix is strictly upper triangular.

The original motivation for $\boldsymbol{T}_N$ is moving the notion of channel polarization in polar codes [3] to the field of rational numbers. For instance, $\boldsymbol{T}_2$ in (2) translates to this field the generation matrix of size 2x2 that is used in polar codes, namely, $\begin{pmatrix} 1 & 1 \\ 1 & 0 \end{pmatrix}$. In fact, the spectral interpretation of polar codes opened the door to the use of various signal processing techniques in polar coding, namely, to exploit the FFT due to the butterfly factor graph of the code. This paper works the other way around and studies how to mimic polar coding procedures and benefit from them in the radio frequency or base band domain. It is a study of the goodness of the so-called radio frequency coding (RFC), as in [6] and [7]. Next we show that combining the proposed transformation, $\boldsymbol{T}_N$, with its corresponding transform pair, $\boldsymbol{T}_N^H$, at reception, the so-called polarized channels are obtained. After that, we propose to recursively partition $\boldsymbol{T}_N$; again, mimicking the original construction of polar codes, which relies on the recursive application of a linear transformation.

Namely, when $\boldsymbol{T}_N^H$ is applied to the received baseband samples in $\boldsymbol{y}$, the resulting equivalent channel is

$$\boldsymbol{T}_N^H \boldsymbol{H}_{ZN} \boldsymbol{T}_N = \begin{pmatrix} \boldsymbol{H}_{N/2} + \boldsymbol{H}_{CN/2} & \boldsymbol{0}_{N/2} \\ \boldsymbol{0}_{N/2} & \boldsymbol{W}_{N/2}^H (\boldsymbol{H}_{N/2} - \boldsymbol{H}_{CN/2}) \boldsymbol{W}_{N/2} \end{pmatrix} =$$

$$= \begin{pmatrix} \boldsymbol{H}_{ZN/2}^+ & \boldsymbol{0}_{N/2} \\ \boldsymbol{0}_{N/2} & \boldsymbol{W}_{N/2}^H \boldsymbol{H}_{ZN/2}^- \boldsymbol{W}_{N/2} \end{pmatrix} \quad (7)$$

Thus, the channel has been split into two polarized channels or physical layer slices. We denote the positive polarized channel as $\boldsymbol{H}_{ZN/2}^+ = \boldsymbol{H}_{N/2} + \boldsymbol{H}_{CN/2}$ and the basic negative polarized channel as $\boldsymbol{H}_{ZN/2}^- = \boldsymbol{H}_{N/2} - \boldsymbol{H}_{CN/2}$. We call polarization or slicing step to the operation in (7), because, as explained in next sections, we can control its associated mutual information (MI) and latency. Thus, differently to the polar codes, the advantage of the so-called polarization transformation is not to correct errors or reduce the bit error rate, but to transform the physical channel into channels with controlled reliability.

On the one hand, matrix $\boldsymbol{H}_{ZN/2}^+$ preserves the circular symmetry and, whenever $L \le N/4$ it can be decomposed again into the previously introduced block matrices, now of size *N/4*

$$H_{ZN/2}^+ = \begin{pmatrix} H_{N/4} & H_{CN/4} \\ H_{CN/4} & H_{N/4} \end{pmatrix}. \qquad (8)$$

Therefore, we could apply again the proposed transformation to this equivalent channel in order to further split it into two polarized channels. As $H_{ZN/2}^+ \in C^{N/2 \times N/2}$ the transformation to be applied now is $T_{N/2}$. This feature can be exploited to recursively slice the OFDM symbol.

On the other hand, the basic negative polarized $H_{ZN/2}^-$ does not preserve the circular symmetry. In fact, its formulation into block matrices is

$$H_{ZN/2}^- = \begin{pmatrix} H_{N/4} & -H_{CN/4} \\ H_{CN/4} & H_{N/4} \end{pmatrix}. \qquad (9)$$

Therefore, as it can be verified, applying the transform pair $\left(T_{N/2}, T_{N/2}^H\right)$ does not result in positive and negative polarized channels. Next section studies the MI of the equivalent transformed channel in (7) and how it splits between the polarized channels.

## III. SPLITTING THE MUTUAL INFORMATION

Since the used transformation is orthonormal, the MI remains the same after its application. In other words, the MI of $H_{ZN}$ is the same as that of $T_N^H H_{ZN} T_N$.

For the purpose of physical layer slicing it is of interest to know the mutual information of each slice, namely, that of the positive and of the negative polarized channels, respectively. Let us consider uniform power allocation, $P$, and formulate the MI of the equivalent channel in (7) as

$$MI = logdet\left(I_N + \frac{P}{\sigma^2} T_N^H H_{ZN} H_{ZN}^H T_N\right) [b/s/Hz]. \quad (10)$$

Due to the block diagonal structure of (7), (10) can be decomposed into

$$MI = MI^+ + MI^-, \qquad (11)$$

where

$$MI^+ = logdet\left(I_{N/2} + \frac{P}{\sigma^2} H_{ZN/2}^+ H_{ZN/2}^{+H}\right) [b/s/Hz], \quad (12)$$

and

$$MI^- = logdet\left(I_{N/2} + \frac{P}{\sigma^2} W_{N/2}^H H_{ZN/2}^- H_{ZN/2}^{-H} W_{N/2}\right) [b/s/Hz]. \quad (13)$$

Appendix A computes each of these expressions, in terms of the reduced $N/4$ subchannels, to show that they are different, and that this difference is negligible if $L \ll \frac{N}{8}$. In this case, the proposed transform allows to split the OFDM symbol into two parts that approximately share half of the capacity each, as Fig. 1 sketches.

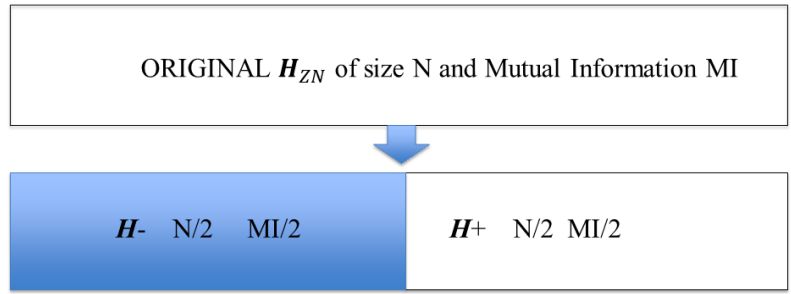

Fig. 1. Each slice shares almost half of the total channel mutual information each if $L \ll \frac{N}{8}$. $H^+$ stands for the positive polarized channel (i.e., $\boldsymbol{H}_{ZN/2}^+$) and $H^-$ for the negative polarized one (i.e., $\boldsymbol{W}_{N/2}^H \boldsymbol{H}_{ZN/2}^- \boldsymbol{W}_{N/2}$).

Therefore, the transmitter can allocate different rates within one symbol by recursively applying the proposed polarization transformation. Most importantly, without CSIT the channel is polarized at transmission into channels with ranked MI. This process is explained in section V and it is key to implement the physical layer slicing. The sharing of the MI is empirically validated in section VI.

## IV. RECEIVER FOR THE POLARIZED CHANNELS

Since the positive polarized channel preserves the circular structure, the receiver is as in OFDM, i.e. FFT and per symbol equalization with the frequency response of the equivalent channel, assumed known at reception. This is not the case for the negative polarized channel, as, in general, it does not preserve circularity. At this point there are two alternatives for its decoding. The first one is direct decoding. The second choice is to use a specific $\boldsymbol{W}_{N/2}$ that facilitates the decoding, which is the one proposed in section IV.b. Although the second alternative is much more effective than the first one, for the sake of completeness, next we present the direct decoding for the negative polarized channel.

### A. Direct decoding of the negative polarized channel

For the sake of clarity, let us consider that In the interest of clarity, let us consider that $\boldsymbol{W}_{N/2} = \boldsymbol{I}_{N/2}$ and that the transformation $\boldsymbol{T}_N$ has only been applied once to an OFDM symbol of $N$ subcarriers (i.e., $N$ temporal samples). The vector $\boldsymbol{s}$ in (3) can be formulated as $\boldsymbol{s} = [\boldsymbol{s}_1 \quad \boldsymbol{s}_2 \quad \boldsymbol{s}_3 \quad \boldsymbol{s}_4]^T$, with $\boldsymbol{s}_i \in C^{N/4 \times 1}$ $i = 1,2,3,4$, and $\boldsymbol{s}_3 \quad \boldsymbol{s}_4$ are the input symbol vectors to the negatively polarized channel. The output vectors $\boldsymbol{z}_3 \quad \boldsymbol{z}_4$ are:

$$\begin{pmatrix} \boldsymbol{z}_3 \\ \boldsymbol{z}_4 \end{pmatrix} = \begin{pmatrix} \boldsymbol{H}_{N/4} \\ \boldsymbol{H}_{CN/4} \end{pmatrix} \boldsymbol{s}_3 + \begin{pmatrix} -\boldsymbol{H}_{CN/4} \\ \boldsymbol{H}_{N/4} \end{pmatrix} \boldsymbol{s}_4. \quad (14)$$

The detection can be successive, i.e. estimates $\boldsymbol{s}_3$ in presence of the interference motivated by $\boldsymbol{s}_4$ and then subtracts the interference of $\boldsymbol{s}_3$ in $\boldsymbol{z}_4$ in order to detect $\boldsymbol{s}_4$. Nevertheless, the complexity of any attempt to detect these two symbols blocks is high. An alternative is to decode the symbols in an iterative manner, by rewriting equation (14) as:

$$\begin{pmatrix} \boldsymbol{H}_{N/4}^{-1} \boldsymbol{z}_3 \\ \boldsymbol{H}_{N/4}^{-1} \boldsymbol{z}_4 \end{pmatrix} = \begin{pmatrix} \boldsymbol{I}_{N/4} & -\boldsymbol{H}_{N/4}^{-1} \boldsymbol{H}_{CN/4} \\ \boldsymbol{H}_{N/4}^{-1} \boldsymbol{H}_{CN/4} & \boldsymbol{I}_{N/4} \end{pmatrix} \begin{pmatrix} \boldsymbol{s}_3 \\ \boldsymbol{s}_4 \end{pmatrix}. \quad (15)$$

This equation suggests the following fixed-point iterative decoding of the negative polarized channel

$$\begin{pmatrix} \boldsymbol{s}_3 \\ \boldsymbol{s}_4 \end{pmatrix}^{m+1} = \begin{pmatrix} \boldsymbol{H}_{N/4}^{-1} \boldsymbol{z}_3 \\ \boldsymbol{H}_{N/4}^{-1} \boldsymbol{z}_4 \end{pmatrix} \begin{pmatrix} \boldsymbol{0}_{N/4} & -\boldsymbol{H}_{N/4}^{-1} \boldsymbol{H}_{CN/4} \\ \boldsymbol{H}_{N/4}^{-1} \boldsymbol{H}_{CN/4} & \boldsymbol{0}_{N/4} \end{pmatrix} \begin{pmatrix} \boldsymbol{s}_3 \\ \boldsymbol{s}_4 \end{pmatrix}^{m} \quad (16)$$

where index $m$ denotes the iteration. In order to perform the iterative decoding, it is necessary to compute basically one matrix inverse and its product with the complement matrix. This is done easily from lower order matrices. In fact, note that from any size below the length of the channel, the inverse verifies the recursive computation for successive orders as indicated in (17)

$$\boldsymbol{H}_n^{-1} = \begin{pmatrix} \boldsymbol{H}_{n-1} & \boldsymbol{0}_{n-1} \\ \boldsymbol{h}^T & \boldsymbol{h}_o \end{pmatrix}^{-1} = \begin{pmatrix} \boldsymbol{H}_{n-1}^{-1} & \boldsymbol{0}_{n-1} \\ -\boldsymbol{h}^T \boldsymbol{H}_{n-1}^{-1} & \boldsymbol{H}_{n-1}^{-1} \end{pmatrix} \quad (17)$$

with $\boldsymbol{h} = [\boldsymbol{h}_n \ \boldsymbol{h}_{n-1} \dots \boldsymbol{h}_1]^T$.

Since a usual size for $N$ in 4G and 5G is 2048 carriers (see [8]), and the tendency of this size it to increase in future radio generations, the described procedure is not effective due to the requirements in memory size and computations. This fact motivates to design $\boldsymbol{W}_{N/2}$ in order to facilitate the decoding of the negative polarized channel.

*B. Decode inspired in the FFT*

Next, we propose $\boldsymbol{W}_{N/2}$ to be diagonal with the following structure

$$\boldsymbol{W}_{N/2} = \begin{pmatrix} \Omega_{N/4} & \boldsymbol{0}_{N/4} \\ \boldsymbol{0}_{N/4} & j\Omega_{N/4} \end{pmatrix}, \quad (18)$$

where $\Omega_{N/4}$ is the complete first step of a time-decimated inverse FFT [9]:

$$\Omega_{N/4} = diag\left(exp\left(j2\pi \frac{m}{N}\right)\right); \quad m = 0, \dots, \frac{N}{4} - 1. \quad (19)$$

If $L \leq N/4$ then the negative polarized channel

$$\boldsymbol{W}_{N/2}^H \boldsymbol{H}_{ZN/2}^- \boldsymbol{W}_{N/2} = \begin{pmatrix} \Omega_{N/4}^H \boldsymbol{H}_{N/4} \Omega_{N/4} & -j\Omega_{N/4}^H \boldsymbol{H}_{CN/4} \Omega_{N/4} \\ -j\Omega_{N/4}^H \boldsymbol{H}_{CN/4} \Omega_{N/4} & \Omega_{N/4}^H \boldsymbol{H}_{N/4} \Omega_{N/4} \end{pmatrix}, \quad (20)$$

is circulant (see appendix B).

The advantage of this new transformation is that the decoding of the resulting negative polarized channel in (21) can be done with the FFT. One can observe that, although the positive channel of size *N/2* is decoded with

$$\frac{N}{2} \log_2 \frac{N}{2} = \frac{N}{2} \log_2 N - \frac{N}{2} \quad (21)$$

operations, the negative channel presents a higher complexity:

$$N + \frac{N}{2} \log_2 \frac{N}{2} = \frac{N}{2} \log_2 N + \frac{N}{2} \quad (22)$$

This is because it requires *N* additional real operations when $\boldsymbol{W}_{N/2}$ is applied at the receiver before the FFT. A higher number of operations for decoding implies a higher latency. In fact, the decoding complexity directly determines the latency that each slice presents. We note that, although the overall decoding complexity (i.e., the sum of (21) and (22)) is the same as that of a FFT of size *N*, $N \log_2 N$, the advantage of the proposed scheme is that we are splitting in a controlled way the MI and the latency in different slices. Figure 2 sketches all the RF coding steps that the information symbols in $\boldsymbol{x}$ undergo when a OFDM symbol is divided into two slices.

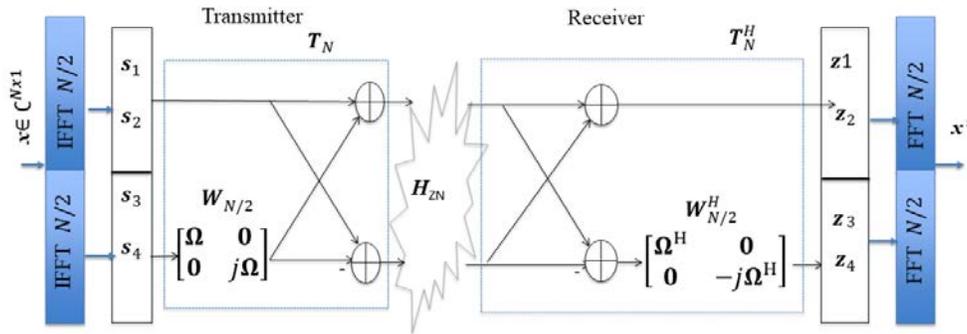

Fig.2 Transmitter and receiver scheme for OFDM transmission with one polarization step in order to create two slices.

After the FFT decoding, each slice must be equalized by the equivalent polarized channel. Remarkably, the estimation of this channel requires the same number of operations as in the usual non-polarized transmission. The reason is twofold: first, the equivalent positive polarized channel in (8) is the same as the original one; second, the equivalent negative polarized channel in (20) is precisely the original channel impulse response modulated at discrete frequency of $\frac{1}{N}$, as appendix B shows. Since in 4G and 5G channel estimation is done based on pilot carriers, next we study the frequency response of the polarized channels.

Let us consider a CIR $h_n$ of size $L \leq \frac{N}{2}$ (i.e., $n = 0, \dots, \frac{N}{2} - 1$), then its DFT of size $N$ is equal to

$$H(l) = \sum_{n=0}^{\frac{N}{2}-1} h_n \, exp\left(-jln\frac{2\pi}{N}\right), \qquad (23)$$

which can be decomposed into its even and odd terms as

$$H(l) = H(2l) + H(2l + 1) \qquad (24)$$

$$H(l) = \sum_{n=0}^{\frac{N}{2}-1} h_n \, exp\left(-jln\frac{2\pi}{N/2}\right) + \sum_{n=0}^{\frac{N}{2}-1} h_n \, exp\left(-j\frac{2\pi}{N}\right) exp\left(-jln\frac{2\pi}{N/2}\right). \qquad (25)$$

Therefore, the frequency response of the equivalent positive and negative polarized channel of size N/2 corresponds to the even and odd terms of $H(l), l = 0, \dots, N-1$, respectively.

## V. GENERAL RECURSIVE ARCHITECTURE

Thanks to the proposed transform, the original channel produces two channels with approximately half of the MI each and different latency or decoding complexity. To obtain these polarized channels the basic operations consist of a channel splitting and a channel combining by using the transform. If the transform is the one proposed in section IV.b, both the positive and the negative polarized channels can be separately decoded as in traditional OFDM. By recursively applying such polarization transform, different slices can be produced that rank in rate and latency; thus, presenting a controlled QoS. In order to specify the details of the recursive transmitter and receiver, without loss of generality, next we consider the case where 4 slices are produced with the design in (18).

First, at the transmitter the stream of information symbols, $x \in C^{Nx1}$ is split into 4 substreams of different size each. Namely, $x = \left[x_{1,\frac{N}{8}} \quad x_{2,\frac{N}{8}} \quad x_{3,\frac{N}{4}} \quad x_{4,\frac{N}{2}}\right]^T$, with $x_{i,p} \in C^{px1}, i = 1,2,3,4; p = \frac{N}{2}, \frac{N}{4}, \frac{N}{8}$. Then the IDFT is carried out accordingly. Equation (26) formulates this process:

$$s = \begin{pmatrix} s_{1,\frac{N}{8}} \\ s_{2,\frac{N}{8}} \\ s_{3,\frac{N}{4}} \\ s_{4,\frac{N}{2}} \end{pmatrix} = diag\left(F_{N/8}^H \quad F_{N/8}^H \quad F_{N/4}^H \quad F_{N/2}^H\right) \begin{pmatrix} x_{1,\frac{N}{8}} \\ x_{2,\frac{N}{8}} \\ x_{3,\frac{N}{4}} \\ x_{4,\frac{N}{2}} \end{pmatrix} \qquad (26)$$

where $F_p^H$ is the IDFT matrix of size $p$. Next, the proposed transformation is applied recursively as (27) indicates:

$$\boldsymbol{x}_T = \boldsymbol{T}_N^R \boldsymbol{s} = \begin{pmatrix} \boldsymbol{T}_{N/2}^R & \boldsymbol{W}_{N/2} \\ \boldsymbol{T}_{N/2}^R & -\boldsymbol{W}_{N/2} \end{pmatrix} \boldsymbol{s} = \begin{pmatrix} \boldsymbol{T}_{N/4} & \boldsymbol{W}_{N/4} & & \boldsymbol{W}_{N/2} \\ \boldsymbol{T}_{N/4} & -\boldsymbol{W}_{N/4} & & \boldsymbol{W}_{N/2} \\ \boldsymbol{T}_{N/4} & \boldsymbol{W}_{N/4} & & -\boldsymbol{W}_{N/2} \\ \boldsymbol{T}_{N/4} & -\boldsymbol{W}_{N/4} & & -\boldsymbol{W}_{N/2} \end{pmatrix} \boldsymbol{s} \quad (27)$$

where $\boldsymbol{T}_N^R$ is obtained from the basic transform in (1) by substituting each identity matrix by a recursive application of the transform. Subsequently, the CP is inserted, and the signal transmitted with an appropriate shaping pulse. Figure 4 depicts the corresponding scheme for this transmitter.

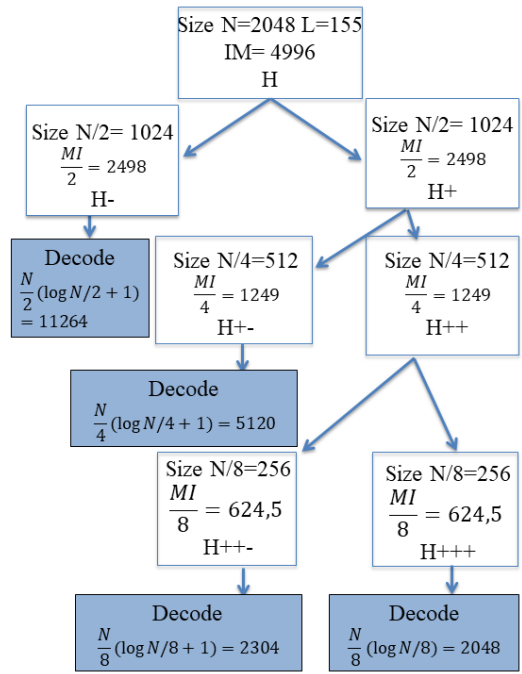

Fig.4. The resulting 4 slices after 3 polarization transforms that are performed on the original channel. The size of the original channel matrix is $N$x$N$ with $N$=2048 and the channel length is $L$=155 and CP=169, which is assumed below $\frac{N}{2^3}$ = 256. MI denotes the mutual information and the negative and positive polarized channels are indicated in the channel matrix. The number of operations is also shown.

At the receiver, once the CP is eliminated, the inverse procedure is carried out by applying the transform pair of $\boldsymbol{T}_N^R$, that is $\boldsymbol{T}_N^{RH}$. As a result, different slices or services are obtained, and each must be equalized by its corresponding equivalent polarized channel. Next, we illustrate this process by using a specific channel of $L$=155 and a transmission of $N$=2048, CP=169. The decoding process can be viewed in Fig.4, where 3 slicing or polarization steps are carried out and 4 slices are produced. The latency is incurred due to the decoding complexity. Figure 4 indicates the complexity at each decoding step. Note that the channel $\boldsymbol{H}_{+++}$ is the one with the lowest latency in delivering the data. The next one is $\boldsymbol{H}_{++-}$ and so forth. Specifically, the number of total operations to decode the 4 slices is 22528, which is the same as the FFT of the overall frame of length $N$=2048 (i.e., $N \log_2 N$ = 22528).

Figure 4 shows a uniform splitting of the MI along the physical slicing process. Remarkably, the splitting can be uniform whenever the structure of the original channel is preserved.

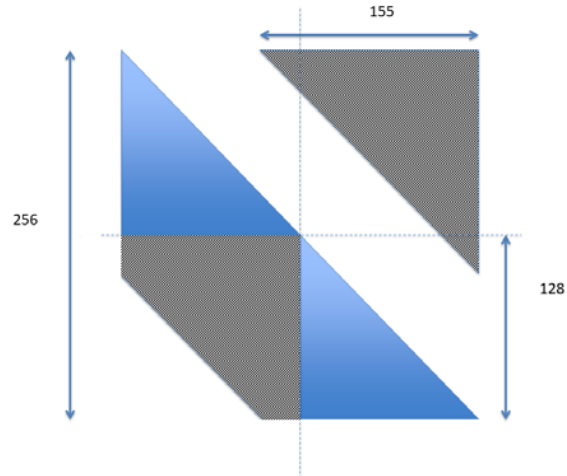

Fig.5. Structure of the last channel matrix $\boldsymbol{H}_{ZM}^{+} \in C^{MxM}$, $M = 256$, which is not valid for further uniform splitting of the MI into two polarized channels of size $\frac{M}{2} = 128$, as $\frac{M}{2}$.

Figure 5 depicts this structure for the example at hand ($N = 2048, L = 155 \leq \frac{N}{2^3} = \frac{N}{8} = 256$): the upper triangular should not overlap with the lower triangular. In other words, $L \leq \frac{N}{2^K}$, where $K$ is the number of total polarization steps or transforms, which produce $K+1$ uniform slices. In Fig. 4, $K+1=3+1=4$. Further splitting can be done, but the produced MI is not uniform any more among the produced slices. Figure 6 plots the rate and latency ranking that results for the whole set of slices. The circle zooms in the zone with the most non-uniform splitting. Note that if the splitting goes beyond the 5th step, then $\frac{N}{2^5} = \frac{2048}{32} = 64$, which is well below the channel length $L$=155. Therefore, the resulting MI cannot be ranked a priory.

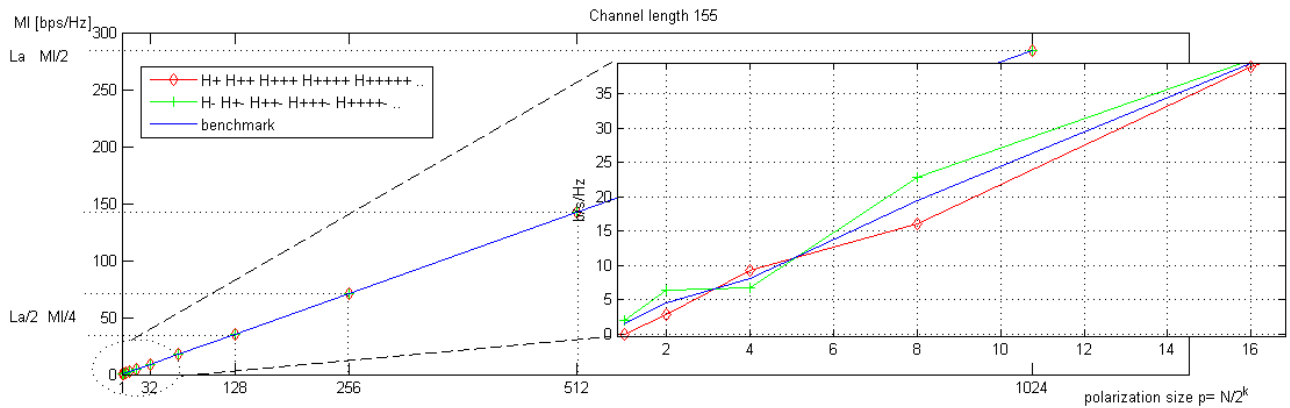

Fig. 6. Mutual Information, MI, and estimated latency, La, for the polarized channels when the splitting goes over, i.e. the last matrix does not obey the structure of (4). $N$=2048, $L$=155.

Table 1 further shows the numerical values of the whole non-uniform zone. Remarkably, as this affects only the low capacity channels, it may not limit, in general, their interest [10].

| size M | 128 | 64 | 32 | 16 | 8 | 4 | 2 | 1 |
|---|---|---|---|---|---|---|---|---|
| $MI_+$ | 314 | 159,2 | 78,77 | 39,93 | 16,5 | 9,24 | 2,914 | 0,043 |
| $MI_-$ | 310,4 | 154,8 | 80,45 | 38,84 | 22,78 | 6,81 | 6,329 | 1,953 |
| Decoding **H** complexity | 1152 | 512 | 224 | 96 | 40 | 16 | 6 | 2/1 |

Table I. MI [b/s/Hz] of Fig. 4 when further non-uniform splitting is carried out from $\mathbf{H}_{+++}$ of size M=256 and with MI=624,5.

## VI. NUMERICAL RESULTS

This section carries out Monte Carlo runs over different 4G/5G channels and system parameters in order to numerically verify the MI ranking along the different slices. The design in section IV.b is used. For a first set of simulations we consider N=2048 and a carrier spacing of $\Delta_f = 15$ kHz, which is mostly used for sub-6 GHz transmissions. The ETU (Extended Typical Urban) channel is considered. Note that the last echo in this channel comes after 5000 ns, as the sampling time is equal to 1/(2048*15000)=32,55 ns, then the channel duration is L=155 taps.

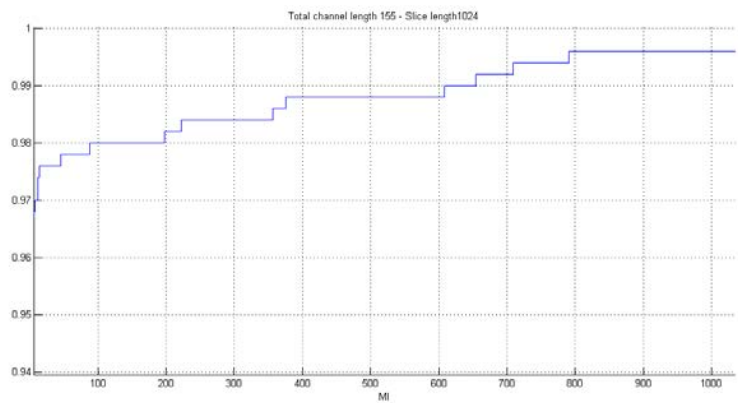

Fig. 7. MI distribution for the first slice of size N/2=1024 and ETU channel of L=155. Three cdf curves are plotted and coincide in the same one: the benchmark, the MI of the positive polarized channel, and the MI of the negative polarized channel.

Figure 7 shows the cumulative density function (cdf) of the MI for 500 channel realizations. Specifically, it plots the MI of both the positive and the negative polarized channels of size N/2=1048 (i.e., first splitting). As benchmark, the ideal MI splitting is considered, that is, when half of MI goes to the positive and to the negative polarized channels, respectively. As expected for this first slice, where $L \ll \frac{N}{2}$, the three plotted cdf curves coincide. In other words, the MI is equally shared between both polarized channels (i.e., H+ and H-).

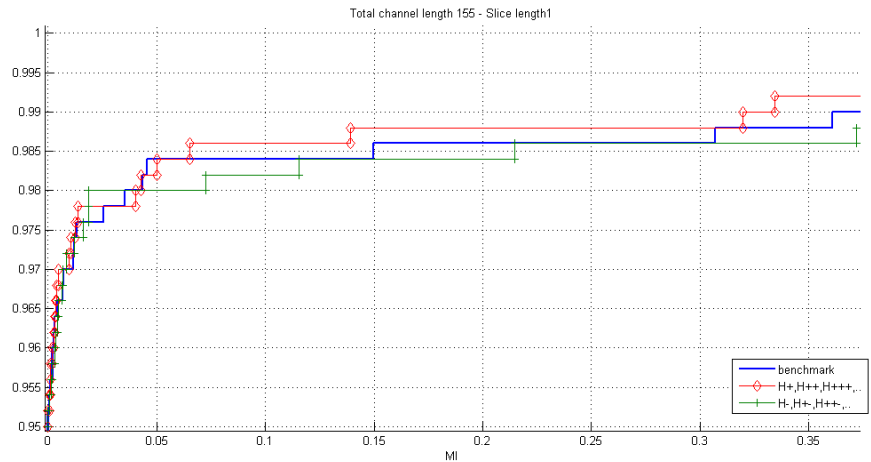

Fig. 8. MI distribution for the last slice of size N/2048=1 and ETU channel of L=155. Three cdf curves are plotted: the benchmark, the MI of the positive polarized channel, and the MI of the negative polarized channel. The curves are slightly different due to the non-uniform splitting.

Next, Fig. 8 depicts the MI distributions for the polarized channels of size N/2048=1 (i.e., last splitting). We remark that in this case the three cdf curves can be distinguished due to the non-uniform splitting for the low rank channels. In any case, the difference is almost negligible. Thus, supporting the good performance of the proposed transform in terms of MI ranking without CSIT.

For the second set of simulations, Fig. 9 considers the EPA (Extended Pedestrian) channel, a carrier spacing of $\Delta_f = 240$ KHz, which is more prone to be used at mmWaves, and N=128 carriers. In the EPA channel, the last echo comes after 410 ns, as the sampling time is 1/(128*240000)=32,55 ns, the channel duration is now shorter, L=14 taps. Figure 9 plots the rate and latency ranking that results for the whole set of slices. The MI is averaged over 50 Monte Carlo runs. From $\frac{N}{2^3} = 16$ (i.e., less than L=14) a non-uniform splitting is expected. The circle amplifies the zone from $\frac{N}{2^5} = 4$. Note again that the non-uniform MI sharing is almost negligible.

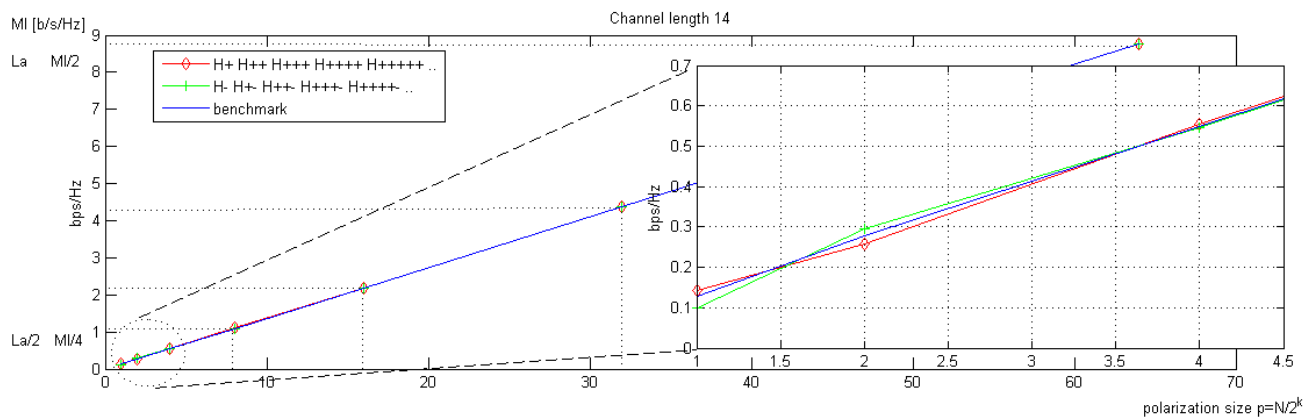

Fig.9. MI slicing for N=128 and EPA channel of L=14. 50 Monte Carlo runs averages the MI at each slice. The circle zooms

in the non-uniform splitting that occurs for the slices of size 4, 2 and 1.

Remarkably, equivalent conclusions have been obtained from simulations with other channels, and different N and $\Delta_f$ combinations.

## VII. CONCLUSIONS AND FUTURE WORK

5G networks rely to a large degree on Software Define Network technologies, which depend very much on the reconfigurability capabilities of the radio equipment and OFDM-based transport. This paper proposes a new type of carrier processing or RFC to meet the user demands in terms of rate and latency. Specifically, we propose an orthonormal transform that operates at OFDM symbol level. The goal is, without the need of CSIT, to polarize the channel into slices with controlled capacity and decoding complexity or latency. In this way, different services can be sliced within the $N$ carriers of an OFDM symbol, all embraced by the same CP and without increasing the complexity of the general OFDM processing.

This work further develops the scalable numerology that 5G presents and just would need of a physical slice format indication to inform the user terminal about the number of slices within the OFDM symbol. At first blush, the proposed slicing at the physical layer might seem a minor step forward from the existing slicing techniques. However, spectrum is a scarce resource and the more it can be fine grained, and the less we over-provision it, the better to unlock new business opportunities. In future work it is of interest the study of its applicability to different virtual radio access network architectures [11]. These architectures can enable, for instance, hybrid terrestrial-satellite multicast, where the same frame is received by multiple users and each one may have a different Service Level Agreement (SLA). Also, of special interest is to study the potential of the proposed technique when applied to promising beyond 5G multicarrier systems that are OFDM-based, as those in [12].

## Appendix A

Let us consider that $L \leq N/4$ so that we can substitute (8) into (12). As a result we obtain that $MI^+$ is $\log \det$ of $\mathbf{I}_{N/2}$ plus the following matrix

$$\begin{pmatrix} \mathbf{H}_{N/4}\mathbf{H}_{N/4}^H + \mathbf{H}_{CN/4}\mathbf{H}_{CN/4}^H & \mathbf{H}_{N/4}\mathbf{H}_{CN/4}^H + \mathbf{H}_{CN/4}\mathbf{H}_{N/4}^H \\ \mathbf{H}_{CN/4}\mathbf{H}_{N/4}^H + \mathbf{H}_{N/4}\mathbf{H}_{CN/4}^H & \mathbf{H}_{N/4}\mathbf{H}_{N/4}^H + \mathbf{H}_{CN/4}\mathbf{H}_{CN/4}^H \end{pmatrix}, \tag{A.1}$$

where $\frac{P}{\sigma^2} = 1$. Operating in a similar way, by substituting (9) into (13), we obtain that $MI^-$ is $\log \det$ of $\mathbf{I}_{N/2}$ plus the following matrix

$$\begin{pmatrix} \mathbf{H}_{N/4}\mathbf{H}_{N/4}^H + \mathbf{H}_{CN/4}\mathbf{H}_{CN/4}^H & \mathbf{H}_{N/4}\mathbf{H}_{CN/4}^H - \mathbf{H}_{CN/4}\mathbf{H}_{N/4}^H \\ \mathbf{H}_{CN/4}\mathbf{H}_{N/4}^H - \mathbf{H}_{N/4}\mathbf{H}_{CN/4}^H & \mathbf{H}_{N/4}\mathbf{H}_{N/4}^H + \mathbf{H}_{CN/4}\mathbf{H}_{CN/4}^H \end{pmatrix}. \tag{A.2}$$

where we have considered $\mathbf{W}_{N/2} = \mathbf{I}_{N/2}$ for this first part of the demonstration.

Due to their different off-diagonal elements, (A.1) and (A.2) have a different determinant. To analyze this difference, next these terms are formulated in detail by further decomposing the channel of size $N/4$:

$$\mathbf{H}_{N/4}\mathbf{H}_{CN/4}^H = \begin{pmatrix} \mathbf{B}_{N/8} & \mathbf{0}_{N/8} \\ \mathbf{D}_{CN/8} & \mathbf{B}_{N/8} \end{pmatrix} \begin{pmatrix} \mathbf{0}_{N/8} & \mathbf{0}_{N/8} \\ \mathbf{D}_{CN/8}^H & \mathbf{0}_{N/8} \end{pmatrix} = \begin{pmatrix} \mathbf{0}_{N/8} & \mathbf{0}_{N/8} \\ \mathbf{B}_{N/8}^H\mathbf{D}_{CN/8}^H & \mathbf{0}_{N/8} \end{pmatrix} \tag{A.3}$$

$$\mathbf{H}_{N/4}\mathbf{H}_{CN/4}^H - \mathbf{H}_{CN/4}\mathbf{H}_{N/4}^H = \begin{pmatrix} \mathbf{0}_{N/8} & -\mathbf{D}_{CN/8}\mathbf{B}_{N/8}^H \\ \mathbf{B}_{N/8}\mathbf{D}_{CN/8}^H & \mathbf{0}_{N/8} \end{pmatrix} \tag{A.4}$$

$$\mathbf{H}_{N/4}\mathbf{H}_{CN/4}^H + \mathbf{H}_{CN/4}\mathbf{H}_{N/4}^H = \begin{pmatrix} \mathbf{0}_{N/8} & \mathbf{D}_{CN/8}\mathbf{B}_{N/8}^H \\ \mathbf{B}_{N/8}\mathbf{D}_{CN/8}^H & \mathbf{0}_{N/8} \end{pmatrix}. \tag{A.5}$$

Note that the off-diagonal terms in (A.4) and (A.5) are small whenever $L << N/8$, and in this case the determinants of the matrices (A.1) and (A.2) can be considered almost equal. This conclusion is valid not only for $\mathbf{W} = \mathbf{I}_{N/2}$ in (13), but for any $\mathbf{W}$ orthonormal, since this transformation preserves the $MI$.

<div align="center">APPENDIX B</div>

*Lemma 1:* Given the symmetric matrix $\mathbf{A}$

$$\mathbf{A} = \begin{pmatrix} \mathbf{\Omega}_{N/4}^H \mathbf{H}_{N/4} \mathbf{\Omega}_{N/4} & -j\mathbf{\Omega}_{N/4}^H \mathbf{H}_{CN/4} \mathbf{\Omega}_{N/4} \\ -j\mathbf{\Omega}_{N/4}^H \mathbf{H}_{CN/4} \mathbf{\Omega}_{N/4} & \mathbf{\Omega}_{N/4}^H \mathbf{H}_{N/4} \mathbf{\Omega}_{N/4} \end{pmatrix}, \tag{B.1}$$

with sub-matrices $\mathbf{H}_{N/4}$ and $\mathbf{H}_{CN/4}$ lower and strictly upper triangular, respectively, and $\mathbf{\Omega}$ that of (18), matrix $\mathbf{A}$ is Toeplitz and circulant.

*Proof:* Note that

$$a_{p,q} = \left[\mathbf{\Omega}_{N/4}^H \mathbf{H}_{N/4} \mathbf{\Omega}_{N/4}\right]_{p,q} = h_{p-q} exp\left(j2\pi(q-p)/N\right),$$

$$p = 1...\frac{N}{4}$$

$$q = 1...p \tag{B.2}$$

$$p - q \leq L$$

$$L < N/4.$$

Clearly, this transformation modulates the impulse response embedded in matrix $\mathbf{H}_{N/4}$ and preserves its Toeptliz structure: $a_{p,q} = a_{(p+1),(q+1)}$. With the same reasoning it can be proved that the off-diagonal sub-matrix $-j\mathbf{\Omega}_{N/4}^H \mathbf{H}_{CN/4} \mathbf{\Omega}_{N/4}$ of $\mathbf{A}$ is Toeplitz.

Also, note that the first row of $-j\boldsymbol{\Omega}_{N/4}^{H}\mathbf{H}_{CN/4}\boldsymbol{\Omega}_{N/4}$ is (B.3):

$$\left[-j\boldsymbol{\Omega}_{N/4}^{H}\mathbf{H}_{CN/4}\boldsymbol{\Omega}_{N/4}\right]_{1,(N/4)-q+1} = -jh_q\,exp\,(j2\pi((N/4)-q)/N) = h_q exp\,(-j2\pi q/N)$$

$$q = 1, ..., L-1 \quad \text{(B.3)}$$

$$L < N/4.$$

Since

$$\left[-j\boldsymbol{\Omega}_{N/4}^{H}\mathbf{H}_{CN/4}\boldsymbol{\Omega}_{N/4}\right]_{1,(N/4)-q+1} = h_q exp\,(-j2\pi q/N) = \left[\boldsymbol{\Omega}_{N/4}^{H}\mathbf{H}_{N/4}\boldsymbol{\Omega}_{N/4}\right]_{q+1,1}$$

$$q = 1, ..., L-1, \quad \text{(B.4)}$$

$$L < N/4.$$

then $\mathbf{A}$ is a circulant matrix. This matrix is derived from a new channel, which is precisely the original channel impulse response modulated at discrete frequency of $1/N$:

$$h_q exp\,(-j2\pi q/N) \qquad q = 1, ..., L-1. \quad \text{(B.5)}$$

∎

ACKNOWLEDGEMENT

This work has received funding from the ministry of Science, Innovation and Universities under project TERESA-TEC2017-90093-C3-1-R (AEI/FEDER,UE) and from the Catalan Government (2017-SGR-1479).

REFERENCES

[1] J. Ordonez-Lucena, O. Adamuz-Hinojosa, P. Ameigeiras, P. Mu˜noz, J. J. Ramos-Mu˜noz, J. F. Chavarria, and D. Lopez,"The creation phase in network slicing: From a service order to an operative network slice," in 2018 European Conference on Networks and Communications (EuCNC), 2018.

[2] S. D'Oro, F. Restuccia, A. Talamonti, and T. Melodia, "The Slice is Served: Enforcing Radio Access Network Slicing in Virtualized 5G Systems," in Proceedings of IEEE International Conference on Computer Communications, April 2019.

[3] E. Arikan, "Channel Polarization: A Method for Constructing Capacity-Achieving Codes for Symmetric Binary-Input Memoryless Channels," IEEE Transactions on Information Theory, vol. 55, no. 7, pp. 3051 –3073, April 2009.

[4] B. Muquet, M. de Courville, and P. Duhamel, "Subspace-based blind and semi-blind channel estimation for ofdm systems," IEEE Transactions on Signal Processing, vol. 50, no. 7, pp. 1699–1712, 2002.

[5] Q. Feng and H. Li, "The research and realization of svd algorithm in ofdm system," in 2010 3rd International Congress on Image and Signal Processing, vol. 9, 2010, pp. 4467–4471.

[6] A. Perez-Neira and M. A. Lagunas, "Radio Frequency Coding," in IEEE APWC IEEE APS topical Conference on antennas and propagation in wireless communications, Granada, Spain, September 2019.

[7]  C. Diaz, A. Perez-Neira, and M. A. Lagunas, "RSBA-Resource Sharing Beamforming access for 5G-mMTC," in IEEE Globecom 2020, Hawai, USA, December 2019.

[8]  E. Dahlman, S. Parkvall, and J. Skold, "5G NR: The Next Generation Wireless Access Technology," in Academic Press, 2018 Elsevier, 2019.

[9]  E. O. Brigham, "The Fast Fourier Transform," in Prentice-Hall, Inc., 1974.

[10] S. S. Lekshmi, M. S. Anjana, B. B. Nair, D. Raj, and S. Ponnekanti, "Framework for generic design of massive iot slice in 5g," in 2019 International Conference on Wireless Communications Signal Processing and Networking (WiSPNET), 2019.

[11] https://www.o ran.org/, "Operator Defined Next Generation RAN Architecture and Interfaces."

[12] H. Kim, Y. Park, J. Kim, and D. Hong, "A low-complex svd-based f-ofdm," IEEE Transactions on Wireless Communications, vol. 19, no. 2, pp. 1373–1385, 2020.